\documentclass[prd,preprint,nofootinbib,amsmath,amssymb,superscriptaddress]{revtex4-1}
\usepackage[dvipdfmx]{graphicx}% Include figure files
\usepackage{dcolumn}% Align table columns on decimal point
\usepackage{bm}% bold math
\usepackage{braket}
\usepackage[colorlinks=true,link color=blue,cite color=blue,urlcolor=black]{hyperref}
\begin{document}

%\preprint{ICRR-Report-662-2013-11, IPMU-13-204}
\vspace*{2cm}
\title{\Large Inflatonic baryogenesis with large tensor mode}% Force line breaks with \\

 \author{Naoyuki Takeda}%
 \email{takedan@icrr.u-tokyo.ac.jp}
\affiliation{%
 Institute for Cosmic Ray Research, University of Tokyo, \\
 Kashiwa, Chiba, 277-8582, Japan}
\date{\today}% It is always \today, today,

\begin{abstract}
We consider the complex inflaton field with a CP asymmetric term for its potential.
This CP asymmetric term produces the global charge of the inflaton after inflation.
With the assignment of the baryon number to the inflaton, the baryon asymmetry of the universe is produced by inflaton's decay.
In addition to this, the $U(1)$ breaking term modulates the curvature of the inflaton radial direction depending on its phase, which affects the tensor to scalar ratio.
In this paper, we have studied the relation between the baryon asymmetry and the tensor to scalar ratio, then verified that the
future CMB observation could test this baryogenesis scenario with large tensor mode.
\end{abstract}

\pacs{Valid PACS appear here}% PACS, the Physics and Astronomy
\maketitle

%\tableofcontents

%%%%% INTRODUCTION%%%%%%%%%%%%%%%%%%%%%%%%%%%%%%%%%%%%%%%%%%%%%%%%%%%%%

\section{Introduction}
%%%%%%%%%%%%%%%%%%%%%%%%%%%%%%%%%%%%%%%%%%%%%%%%%%%%%%%%%%%%%%%%%%%
Inflation theory is attractive to solve the cosmological problems such as the flatness, horizon and monopole problems.
Furthermore, inflaton's quantum fluctuations give the seeds of the large scale structures, and they are imprinted on the cosmic microwave background (CMB).
Observations of the CMB such as the Planck mission~\cite{Ade:2015lrj}, have verified that the spectral index of the scalar perturbation deviates from the flat spectrum with more than $5\sigma$, which 
strongly suggests the realization of the inflation. 
By the detection of the tensor mode in the near future, this inflation theory and its energy scale will be confirmed.

The domination of the inflaon's energy over the universe induces the cosmic accelerating expansion, which dilutes the other matter contents and their asymmetry presented before inflation.
On the other hand, by the observation of CMB~\cite{Ade:2015xua} and by the measurements of the primordial abundances of the light elements~\cite{Cooke:2013cba},
the abundance of the baryon asymmetry after the Big Bang Nucleosynthesis is confirmed as $n_s/s\simeq10^{-10}$.
To explain this asymmetry, various production mechanisms have been proposed such 
the electroweak baryogenesis~\cite{Kuzmin:1985mm}, the leptogenesis~\cite{Fukugita:1986hr} 
and the Affleck-Dine baryogenesis~\cite{Affleck:1984fy}.
In this paper, we focus on the Affleck-Dine baryogenesis, in which the rotating complex scalar field  (AD field) on the field space with baryon number produces the $U(1)_B$ charge,
and then by the decay of the AD field into baryon, the baryon asymmetry is produced. 
We assign the baryon number on the inflaton, and investigate the baryon production by the inflaton's AD mechanism.

Supposed the quadratic chaotic inflation~\cite{Linde:1983gd}, the constraint for the abundance of the tensor mode by Planck mission~\cite{Ade:2015lrj} suggests that
the potential takes some suppression by the higher terms such cubic one as $V\sim m^2\phi^2-\lambda\phi^3$.
In the previous studies~\cite{Hertzberg:2013jba,Hertzberg:2013mba,Charng:2008ke}, it is showed that if this higher term breaks $U(1)$ symmetry such as $V=m^2|\Phi|^2+\lambda\Phi^n+h.c.$,
its breaking gives the variation of the inflaton for the phase direction after the inflation and then the asymmetry of the inflaton is produced.
Furthermore, \cite{Hertzberg:2013jba,Hertzberg:2013mba,Charng:2008ke} have showed that by the decay of the inflaton assigned baryon number, the large amount of the 
baryon asymmetry is produced to explain the observed one $n_s/s\sim10^{-10}$.

As showed in~\cite{Hertzberg:2013jba,Hertzberg:2013mba}, the abundance of the tensor mode for this model is mainly determined by the quadratic term $V\sim m^2|\Phi|^2$, however, 
the higher terms $V\ni\lambda\Phi^n$ would also gives the sizable modulation of the the tensor mode.
This higher term is also the source of the inflaton's asymmetry, which thus means that the tensor mode is correlated with the inflaton's asymmetry, and then with the baryon asymmetry. 
In this paper, we have investigated the relation between the prediction for the tensor mode and the abundance of the baryon asymmetry produced by the inflaton,
supposing the polynomial inflation where $U(1)$ symmetry is broken by the cubic term as $V=m^2|\Phi|^2+(\lambda\Phi^3+h.c.)+\lambda^2/m^2|\Phi|^4$. 

The organization of this paper is as follows. 
At first in Sec.~\ref{sec:inflation dynamics}., we briefly explain the dynamics of this inflation model, and then calculate the prediction for tensor to scalar ratio $r$ and
spectral index $n_s$ for each initial phase of the inflaton and for the typical strength of the coupling $\lambda$.
Then, in Sec.~\ref{sec:baryon production}, we calculate the inflaton's asymmetry produced in this model, and then discuss its decay into baryons.
Finally, we conclude this paper in Sec.~\ref{sec:conclusion}.

%%%%% inflation dynamics%%%%%%%%%%%%%%%%%%%%%%%%%%%%%%%%%%%%%%%%%%%%%%%%%%%%%

\section{inflation dynamics}
\label{sec:inflation dynamics}
%%%%%%%%%%%%%%%%%%%%%%%%%%%%%%%%%%%%%%%%%%%%%%%%%%%%%%%%%%%%%%%%%%%
In this section, we show the model and briefly explain its dynamics. Then, we investigate the prediction for CMB observations $r$ and $n_s$. 

We consider the complex scalar inflaton $\Phi$, whose action is given as
%%%%%
\begin{equation}
\label{eq_id_1}
S=\int d^4x\sqrt{-{\rm det}(g_{\mu\nu})}\left[\frac{M_p^2}{2}R+|\partial_{\mu}\Phi|^2-V(\Phi)\right],
\end{equation}
where $R$ is the Ricci scalar.
We give the inflaton's potential within the renormalizability as
%%%%%
\begin{equation}
\label{eq_id_2}
V=m^2|\Phi|^2+\lambda(\Phi^3+\Phi^{\ast 3})+g|\Phi|^4,
\end{equation}
where $m$ is the mass, $\lambda$ is the dimension one constant, and $g$ is the dimension less coupling,
and we have supposed a CP asymmetric cubic term in the potential.
For the simplicity, we set the coupling $g$ by the dimension one parameter $\lambda$ as $g=\lambda^2/m^2$.
\footnote{
This relation between $\lambda$ and $g$ as $g=\lambda^2/m^2$ is established for the supersymmetric polynomial inflation model~\cite{Nakayama:2013txa}.
}
Separating variable of the inflaton into the radial and phase parts as $\Phi=(\phi/\sqrt{2})\exp(i\theta)$, we can reduce the potential (\ref{eq_id_2}) as
\begin{equation}
V=m^2\left[\frac{\phi^2}{2}+\sqrt{2}\alpha\cos(3\theta)\frac{\phi^3}{M_p}+\alpha^2\frac{\phi^4}{M_p^2}\right],
\end{equation}
where we have defined the dimensionless parameter as $\alpha\equiv\lambda M_p/(2m^2)$.
For this potential, the radial part $\phi$ follows bellow equation as
\begin{equation}
\ddot{\phi}+3H\dot{\phi}+m^2\left[1+3\sqrt{2}\alpha\cos(3\theta)\frac{\phi}{M_p}+4\alpha^2\frac{\phi^2}{M_p^2}\right]\phi=0,
\end{equation}
where the over dot means the cosmic time time derivative and $H$ means the Hubble parameter and we have neglected the spatial derivative of the inflaton. 
In the case of the quadratic chaotic inflation model $V_{\rm chao}=(m^2/2)\phi^2$, the field value at $60$ e-folds number is $\sqrt{60}M_p$.
We suppose that the strength of the asymmetric term is small so that the modulation of the cubic term for the quadratic term is small at $60$ e-folds as $\alpha<10^{-2}$.
In this section, we neglect the dynamical variation of the inflaton's phase during inflation, and set the phase by the constant one as $\theta\simeq\theta_i$.
Then, the Hubble parameter is approximated by the homogeneous mode of the inflaton's radial direction as
$H\simeq\sqrt{(1/3M_p^2)\left[(1/2)\dot{\phi}^2+V\right]}$.

The curvature perturbations produced by inflaton and the graviton's fluctuations are imprinted on CMB as the scalar and  the tensor mode.
The scale dependence of the scalar mode is given by the spectral index $n_s$, and the abundance of the tensor mode is given by the tensor to scalar ratio $r$.
Taking the slow roll approximation, we can give $n_s$ and $r$ by the slow roll parameters $\epsilon$ and $\eta$ as
%%%%%
\begin{eqnarray}
\label{eq_id_6}
n_s&=&1+2\eta-6\epsilon,\\
\label{eq_id_7}
r&=&16\epsilon,
\end{eqnarray}
where $\epsilon\equiv M_p^2/2\left(V_{\phi}/V\right)^2$, and $\eta\equiv M_p^2 V_{\phi\phi}/V$.
From these definitions, we can see  that $n_s$ and $r$ depend on the gradient or the curvature of the inflaton's potential, which are determined by
the coupling of the $CP$ asymmetric term $\alpha$ and inflaton's initial phase $\theta_i$.
Thus, by the determination of $n_s$ and $r$, we can constrain $\alpha$ and $\theta_i$.
Remaining of this section, we calculate the $n_s$ and $r$ at the pivot scale, where we suppose $60$ e-folding number.
Then, compare the prediction by the present constraint by Planck mission~\cite{Ade:2015lrj}.

\begin{figure}[htbp]
\begin{center}
\begin{tabular}{c c}
\resizebox{80mm}{!}{\includegraphics{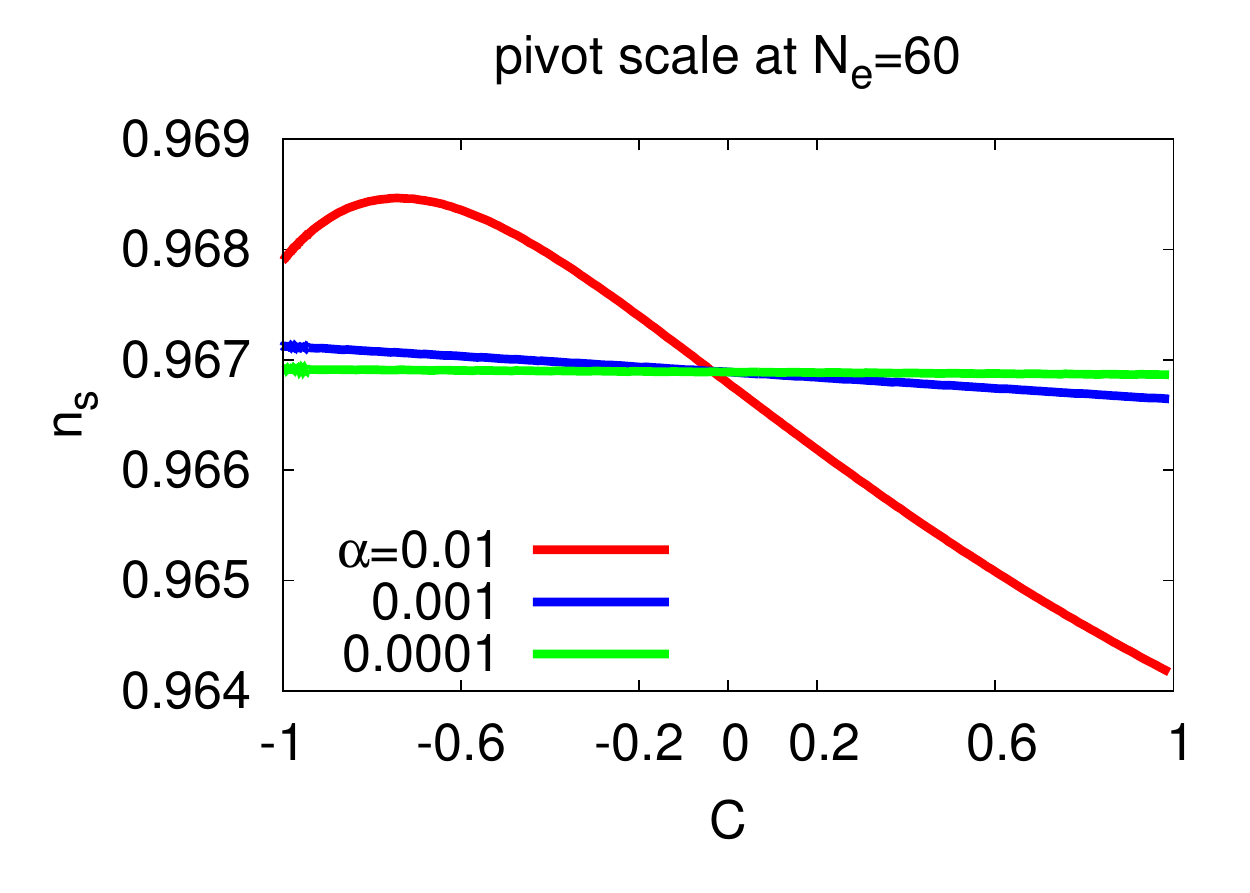}} &
\resizebox{80mm}{!}{\includegraphics{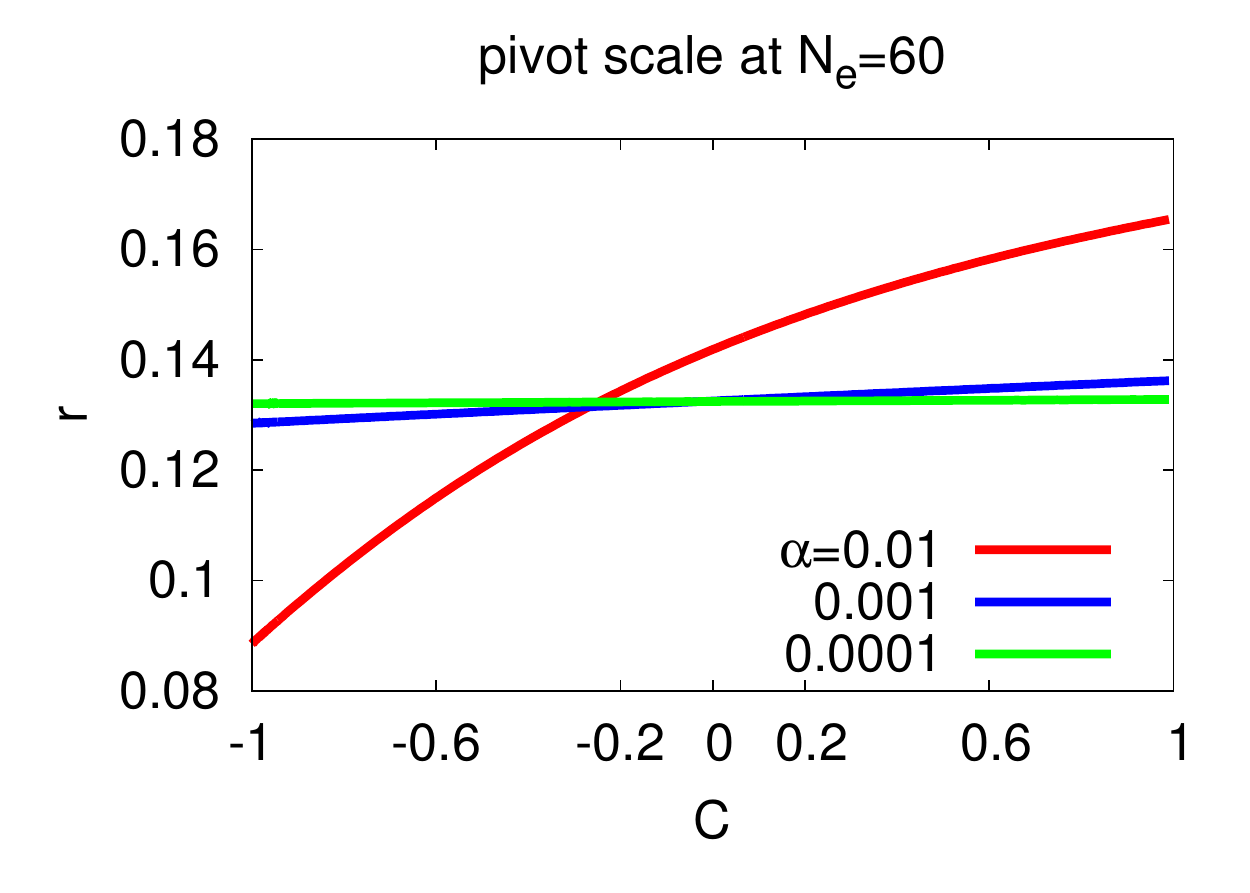}}\\
\resizebox{80mm}{!}{\includegraphics{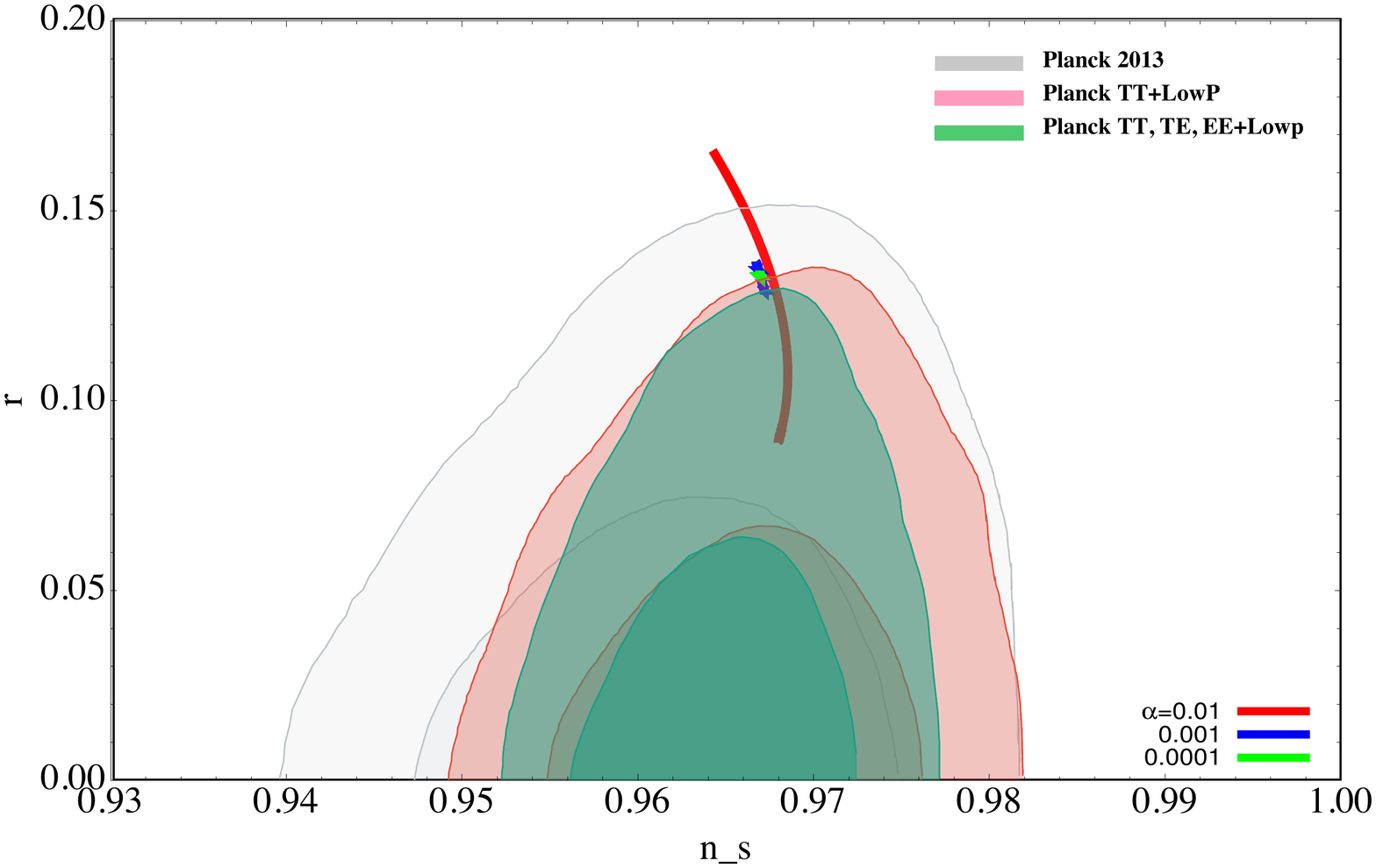}} &
\end{tabular}
\caption{
Plots of $n_s$ and $r$ for each $\alpha$ and $C$, where $C$ is defined by the initial phase of inflaton as $C\equiv\cos(3\theta_i)$.
We have set the parameter $\alpha=10^{-2},10^{-3},10^{-4}$ and $C=-1~{\rm to}~1$.
We set $60$ e-folding number as $N_e=60$.
The top left (right) panel shows the dependence of $n_s~(r)$ on $C$ for each $\alpha$.
The bottom panel shows dependence of the $r-n_s$ relation on $C$ for each $\alpha$ plotted in the observational constraint by Planck~\cite{Ade:2015lrj}.
The thins (thick) region is the one (two) sigma deviation region.
}
\label{FIG_1}
\end{center}
\end{figure}

The e-folding number from the pivot scale $t_i$ to the end of the inflation $t_e$ is given as
%%%%%
\begin{equation}
\label{eq_id_8}
N_e=\int_{t_i}^{t_e}Hdt\simeq\int^{\phi_i}_{\phi_e}\frac{1}{M_p^2}\frac{V}{V_{\phi}}d\phi,
\end{equation}
where $\phi_e (\phi_i)$ is the field value of $\phi$ at $t_e (t_i)$, and we have used the slow roll approximation.
$\phi_e$ is given by the one at the breaking of the slow roll condition as
%%%%%
\begin{equation}
\label{eq_id_9}
{\rm max}\{\epsilon,\eta\}=1.
\end{equation}
In this paper we numerically solve the two equations (\ref{eq_id_8}) and (\ref{eq_id_9}) for $N_e=60$, then substituting the obtained value of $\phi_i$ 
into the equation (\ref{eq_id_6}) and (\ref{eq_id_7}), we evaluate the predictions of $n_s$ and $r$ at the pivot scale.
The results of the simulations are summarized in Fig.~\ref{FIG_1}.
%%%%%%%%%%%%%%%%%%%%%%%%%%%% FIGURE 1 %%%%%%%%%%%%%%%%%%%%%%%%%%%%%%%%%%%%%
%%%%%%%%%%%%%%%%%%%%%%%%%%%%%%%%%%%%%%%%%%%%%%%%%%%%%%%%%%%%%%%%%%%%%%%%%%%
From the top left panel of Fig.~\ref{FIG_1}, we can see that the spectral index depends on initial phase of the inflaton referred as $C\equiv\cos(3\theta_i)$ and its modulation is larger for more larger $\alpha$.
For the negative value of $C$, the potential is suppressed by the cubic term, then the tensor to scalar ratio  becomes smaller showed on the top right panel of Fig.~\ref{FIG_1}.
Our model converges to the quadratic chaotic inflation model for the limit of the small coupling $\alpha\rightarrow 0$~\cite{Linde:1983gd}.
We can see this behavior for $n_s-r$ relation on the bottom panel of Fig.~\ref{FIG_1}.
For the smaller couplings $\alpha=10^{-3},10^{-4}$, the $n_s-r$ relation is the point like one regardless of the initial phase, which is the same pint as the chaotic inflation predicts.
However, for the larger one $\alpha=10^{-2}$, due to the cubic term modulation, the prediction for the tensor to scalar ratio depends on the initial phase of the inflaton,
\footnote{
We have checked the effect of the phase direction's dynamical variation on the e-folding number by the numerical simulation.
Even including the dynamical variation of the phase direction, the deviation of the total e-folding number is smaller than one.
}
 which 
will be tested in the future observation of the CMB.
This initial phase of the inflaton also relate to the amount of the inflaton's asymmetry, then finally relates to the baryon asymmetry, which is described in the next section.

%%%%% baryon production%%%%%%%%%%%%%%%%%%%%%%%%%%%%%%%%%%%%%%%%%%%%%%%%%%%%%
\section{Inflaton asymmetry and its decay into Baryon}
\label{sec:baryon production}
%%%%%%%%%%%%%%%%%%%%%%%%%%%%%%%%%%%%%%%%%%%%%%%%%%%%%%%%%%%%%%%%%%%
During the inflation, the variation for the phase direction is negligible due to the Hubble friction, and after the inflation it stars to roll down to
the minimum $\theta_{\rm min}=n\pi/3,~(n\in{\cal N})$.
This rotation for the phase direction produces $U(1)$ charge of the inflaton. 
At the same time, the Hubble expansion decreases the amplitude of the radial direction, which suppresses the cubic term and then the $U(1)$ charge becomes time independent.
By the decay of inflaton into other particles, the $U(1)$ charge is transferred to the baryon asymmetry. 
In this section, we evaluate the $U(1)$ charge by numerically solving the equation of motion including the dynamical variation of the phase direction. 

The equation of motion of the complex inflaton field $\Phi$ is given as
\begin{equation}
\label{eq_bp_1}
\ddot{\Phi}+3H\dot{\Phi}+\frac{\partial V}{\partial\Phi^{\ast}}=0,
\end{equation}
where the derivative of the potential is given as
\begin{equation}
\frac{\partial V}{\partial\Phi^{\ast}}=m^2\left[\Phi+6\alpha\frac{\Phi^{\ast2}}{M_p}+8\alpha^2\frac{|\Phi|^2}{M_p^2}\Phi\right].
\end{equation}
To integrate this equation, we separate the field into real and imaginary part as
\begin{equation}
\Phi\equiv\frac{\phi_1+i\phi_2}{\sqrt{2}}.
\end{equation}
In simulations, we give the Hubble parameter by the Friedmann equation as
\begin{equation}
H=\sqrt{\frac{\rho_{\Phi}}{3M_p^2}},
\end{equation}
where $\rho_\Phi$ is the energy density of the inflaton defined as
\begin{equation}
\rho_\Phi=|\partial_{\mu}\Phi|^2+V(\Phi)\simeq|\dot{\Phi}|^2+V(\Phi)
=\frac{1}{2}\left[(\partial_\mu\phi_1)^2+(\partial_\mu\phi_1)^2\right]+V(\phi_1,\phi_2),
\end{equation}
where we have neglected the gradient energy of the inflaton.
Using the data of the simulations, we evaluate the $U(1)$ charge of the inflaton asymmetry $\Delta n_\phi$ defined as
\begin{equation}
\label{eq_bp_2}
\Delta n_{\phi}=-i(\Phi^{\ast}\dot{\Phi}-\dot{\Phi}^{\ast}\Phi)=\phi_1\dot{\phi_2}-\phi_2\dot{\phi_1}.
\end{equation}
For the convenience of the bellow calculation, we normalize this asymmetry by the inflaton's number as
\begin{equation}\label{eq_bp_3}
A\equiv\frac{\Delta n_\phi}{n_\phi}=\frac{m}{\rho_{\phi}}\Delta n_{\phi}.
\end{equation}
After the sufficient damping of the inflaton's amplitude, $\Delta n_{\phi}$ is diluted such as the ordinary number density of the matter as $\Delta n_{\phi}\propto a^{-3}$.
Thus, the normalized inflaton asymmetry $A$ converges to the constant value after the sufficient oscillations.
We calculate this constant value of $A$ for each initial phase in $\alpha=10^{-2},10^{-3},10^{-4}$ cases.

\begin{figure}[htbp]
\includegraphics{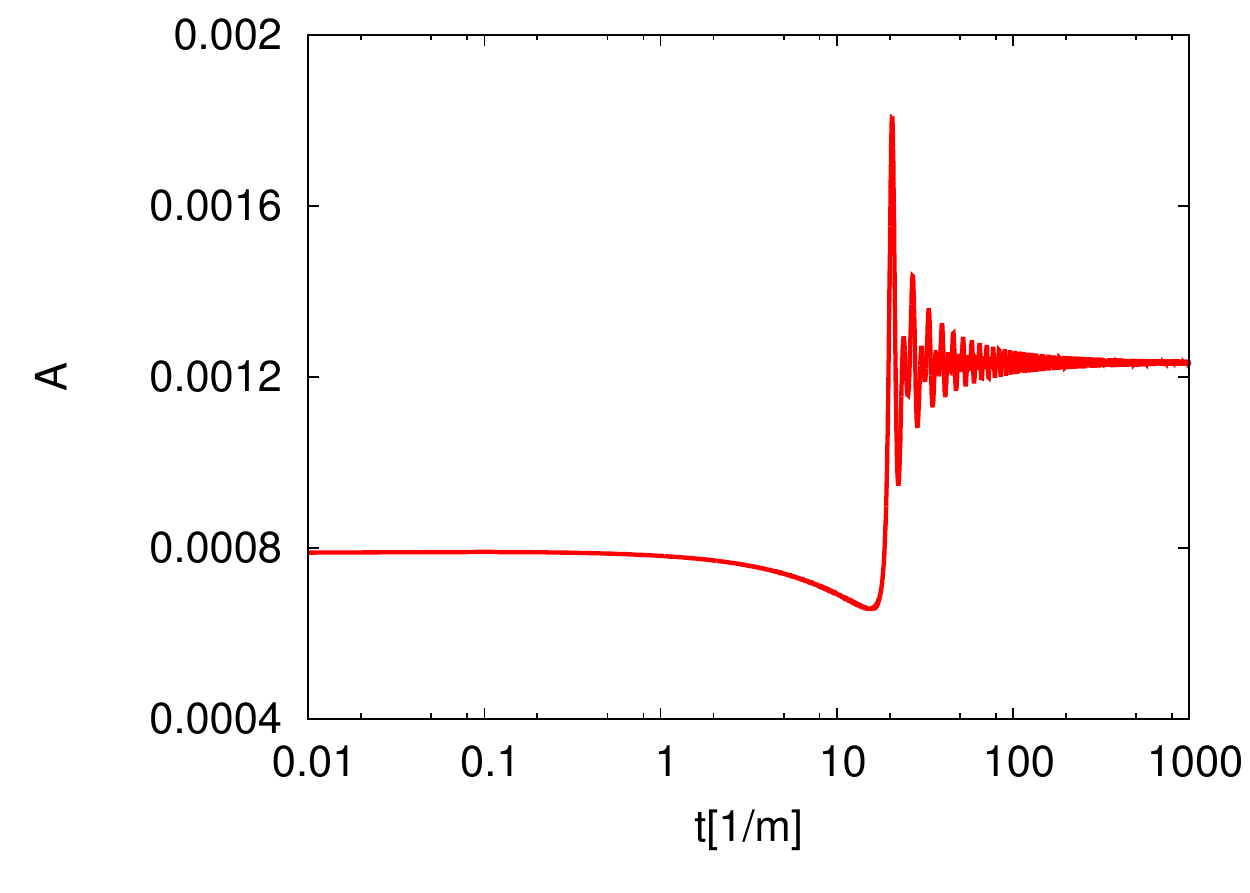}
\caption{
Time evolution of the inflaton asymmetry $A$.
The initial condition of the inflaton is given as $\phi_i=15.4,~\cos(3\theta_i)=-0.994$.
The coupling of the cubic term is given as $\alpha=0.001$.
In this figure, inflation ends at $t\sim20[1/m]$.
}
\label{FIG_2}
\end{figure}

\begin{figure}[htbp]
\includegraphics{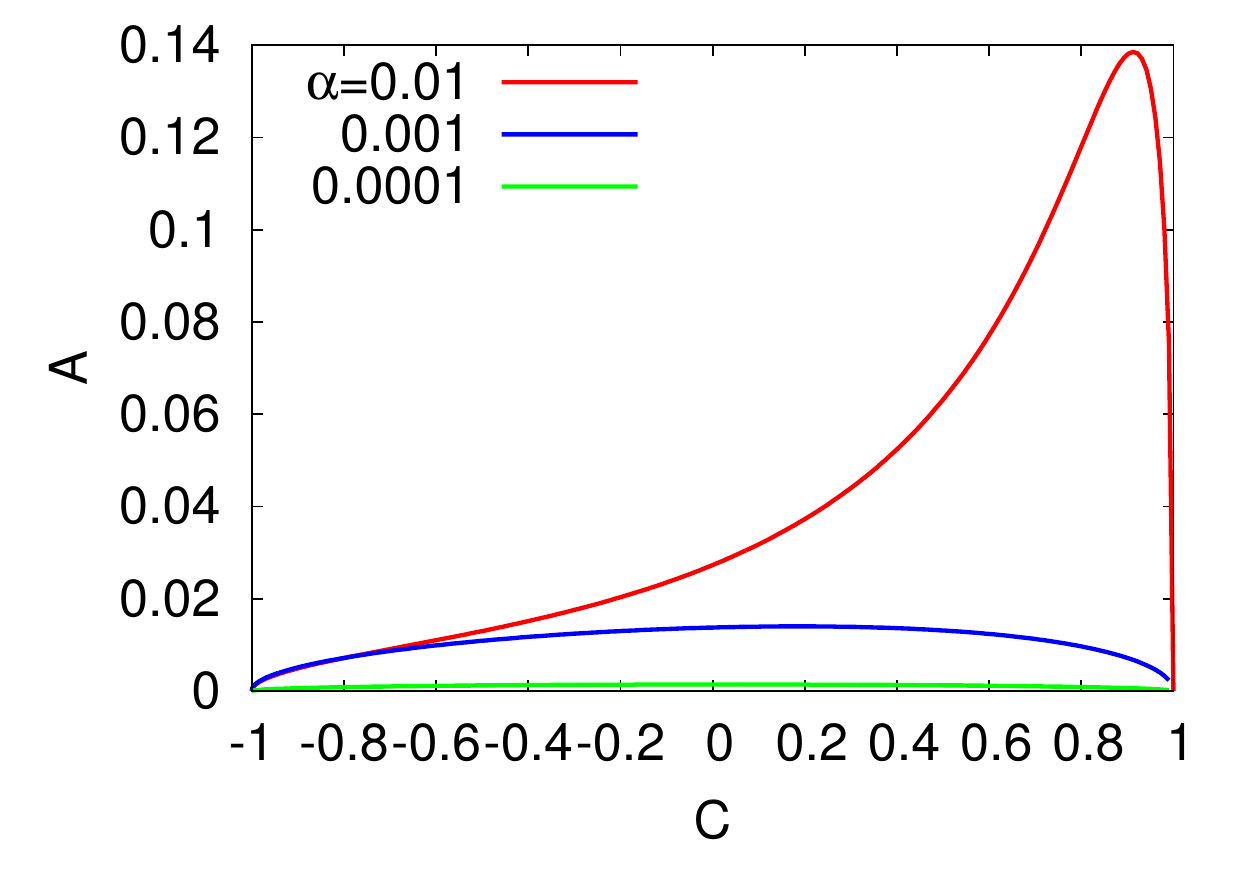}
\caption{
The phase dependence of the inflaton asymmetry $A$ for $\alpha=10^{-2},10^{-3},10^{-4}$.
$C$ is defined by the initial phase as $C\equiv\cos(3\theta_i)$.
}
\label{FIG_3}
\end{figure}

In order to taking account the dynamical variation of the phase direction even during the inflation, we start the simulation from the pivot scale.
We have confirmed that the deviation of the e-folding number from the one neglecting the phase's variation is within $1$ e-folding number.
Simulations are continued until the inflaton's asymmetry $A$ converges to the constant value.
We show the result for one parameter set as $\alpha=10^{-3},\phi_i=15.4,\cos(3\theta_i)=-0.994$ in Fig.~\ref{FIG_2}.
%%%%%%%%%%%%%%%%%%%%%%%%%%%% FIGURE 2 %%%%%%%%%%%%%%%%%%%%%%%%%%%%%%%%%%%%%
%%%%%%%%%%%%%%%%%%%%%%%%%%%%%%%%%%%%%%%%%%%%%%%%%%%%%%%%%%%%%%%%%%%%%%%%%%%
Until the end of the inflation $t\simeq20/m$, inflaton slowly rolls down for the radial and phase directions, thus the asymmetry $A$ is small and nearly constant.
After the end of the inflation, inflaton starts to oscillation and then, by the suppression of the cubic term, $A$ becomes time independent.
This eventual asymmetry $A$ is related with the initial phase of the inflaton $\theta$ and coupling $\alpha$, which is showed in Fi.g~\ref{FIG_3}.
%%%%%%%%%%%%%%%%%%%%%%%%%%%% FIGURE 3 %%%%%%%%%%%%%%%%%%%%%%%%%%%%%%%%%%%%%
%%%%%%%%%%%%%%%%%%%%%%%%%%%%%%%%%%%%%%%%%%%%%%%%%%%%%%%%%%%%%%%%%%%%%%%%%%%
From this figure, we can see that for $\alpha=10^{-2}$, the typical value of the inflaton asymmetry $A$ is $A\sim 10^{-2}$ up to $0.14$.

We suppose that the inflaton has the baryon number $b_{\phi}$.
In this case, the $U(1)$ charge is the baryon asymmetry.
By the decay of the inflaton, this asymmetry $A$ is transferred to the baryon particles.
Under the assumption that the inflaton rapidly decays into radiations when the decay rate $\Gamma$ becomes comparable with the Hubble dilution rate $3H$ (reheating),
the baryon to entropy ratio today $\Delta n_b/s(t_0)$ is estimated by the one at the reheating as
\begin{equation}\label{eq:16}
\begin{split}
\frac{\Delta n_b}{s}(t_0)&\simeq\frac{\rho_{\phi}}{s}(t_{\rm reh})b_{\phi}\frac{A}{m}=
\frac{3}{4}T_Rb_{\phi}\frac{A}{m}\\
&=7.5\times10^{-11}b_{\phi}\frac{A}{0.01}\frac{T_R}{10^5{\rm GeV}}\frac{10^{13}{\rm GeV}}{m},
\end{split}
\end{equation}
where $T_R$ is the reheating temperature. 
From eq.~(\ref{eq:16}), we can see that the baryon to entropy ratio linearly depends on the reheating temperature $T_R$, which is determined by the temperature of the radiation 
at $\Gamma\simeq 3H$ as
\begin{equation}
T_R\simeq(\pi^2 g_{\ast})^{-1/2}\sqrt{\Gamma M_p}\simeq0.1\sqrt{\Gamma M_p},
\end{equation}
where $g_{\ast}$ is the relativistic freedom, and we have supposed as $g_{\ast}\simeq100$.
Thus, for the low decay rate of the inflaton so that the reheating temperature is low as $10^{5}{\rm GeV}$, the 
baryon asymmetry of the Universe is produced by the inflaton's CP asymmetric term.
For the protection of the flatness of the inflaton potential, this low decay rate for the baryonic charged inflaton is plausible such as the dimension five operator suppressed by the GUT scale.
\footnote{
As one realization of the baryonic inflaton's interaction, we mention the higher dimensional super potential based on SUSY as
$W=y\phi\bar{u}\bar{u}\bar{d}/M$, where $y$ is the coupling constant, and $\bar{u},\bar{d}$ are super field of the up and down quarks.
In this case, the decay rate of the inflaton is given as $\Gamma\simeq y^2m^3/M$.
Supposing that the strength of the coupling is gauge's one as $y\simeq10^{-2}$, and supposing that the suppression scale is GUT scale as $M\simeq 10^{16}{\rm GeV}$ for $m=10^{13}{\rm GeV}$, we
obtain the reheating temperature as $T_R\simeq y(M_p/M)10^{9}{\rm GeV}\simeq10^5{\rm GeV}$.%, which explain the baryon asymmetry of the Universe.
}
%%%%%%%%% CONCLUSION %%%%%%%%%%%%%%%%%%%%%%%%%%%%%%%%%%%%%%%%%%%%%%%

\section{CONCLUSION}
\label{sec:conclusion}

%%%%%%%%%%%%%%%%%%%%%%%%%%%%%%%%%%%%
In this paper, we have considered a complex inflaton with the baryon number, whose baryon $U(1)$ asymmetry is slightly broken by the cubic term in the potential.
Depending on the initial phase of the inflaton, the curvature of the inflaton's potential is modulated by the cubic term, and then the tensor to scalar ratio $r$ at the CMB scale
takes the specific value for each initial phase.
On the other hand, the cubic term gives the initial velocity for the phase direction and then by the Affleck Dine mechanism, 
the baryon charge $A$ is produced depending on the initial phase.
After the decay of the inflaton, the inflaton's baryon charge is transferred into the baryon particles.

In this paper, we have evaluated $r$ and $A$, numerically solving the evolution of the inflaton for each initial phase as showed in 
Fig.~\ref{FIG_1} and in Fig.~\ref{FIG_3}.
From the results, we have seen that for the coupling of the cubic term $\alpha$ as large as $10^{-2}$, the 
large asymmetry of the inflaton $A\simeq10^{-2}$ is produced.
For the case that the inflaton has the baryon number, its decay would be induced by the higher dimensional operator, in order to protect the flatness of the inflaton's potential.
As noted in the end of the Sec.~\ref{sec:baryon production}, if this decay is induced by dimension 5 operator whose suppression scale is GUT scale, the reheating temperature becomes 
$T_{R}\simeq 10^{5}{\rm GeV}$, which explains present baryon to entropy ratio (\ref{eq:16}).
For this large coupling $\alpha\simeq10^{-2}$,
the variation of the tensor mode is large as $\Delta r\simeq0.01$ around $r\simeq 0.1$, which is out of $1{\sigma}$ deviation but within $2{\sigma}$ deviation from the constraint of the Planck mission~\cite{Ade:2015lrj}.
Thus, by the future precise observation of the CMB especially tensor mode such as the LiteBIRD mission~\cite{LiteBIRD},
we can verify the relation between $r$ and inflaton's asymmetry $A$, and then the baryon asymmetry for this scenario.

\section*{Acknowledgments}
N.T. thanks Masahiro Kawasaki and Keisuke Harigaya for important discussions.
%%%%%%%%%%%%%%%%%%%%%%%%%%%%%%%%%%%%
%%%%%%%%%%%%%%%%%%%%%%%%%%%%%%%%%%%%
%%%%%%%%%%%%%%%%%%%%%%%%%%%%%%%%%%%%
%%%%%%%%%%%%%%%%%%%%%%%%%%%%%%%%%%%%
%%%%%%%%%%%%%%%%%%%%%%%%%%%%%%%%%%%%

\end{document}